# The black hole final state for the Dirac fields In Schwarzschild spacetime

D. Ahn[1]*, Y. H. Moon[1], R. B. Mann[2]*, and I. Fuentes-Schuller[3]

[1]*Institute of Quantum Information Processing and Systems, University of Seoul, Seoul 130-743, Republic of Korea*

[2]*Department of Physics, University of Waterloo, Waterloo, Ontario, Canada N2L, 3G1*

[3]*Instituto de Ciencias Nucleares, Universidad Nacional Autonoma de Mexico, A-postal 70-543, Mexico D. F. 04510, Mexico*

*Abstract:* We show that the internal stationary state of a black hole for massless Dirac fields can be represented by an entangled state of collapsing matter and infalling Hawking radiation. This implies that the Horowitz-Maldacena conjecture for the black hole final state originally proposed for the massless scalar fields is also applicable to fermionic fields as well. For an initially mixed state we find that the measure of mixedness is expected to decrease under evaporation.

*Corresponding Authors*

*E-mail: dahn@uos.ac.kr ; davidahn@hitel.net; rmann@perimeterinstitute.ca*

The black hole information paradox [1,2] has posed a serious challenge to fundamental physics for over three decades. Hawking's semi-classical argument predicts that the process of black hole formation and evaporation is not unitary [3]. On the other hand, there is evidence from string theory that the formation and evaporation of black holes should be consistent with the basic principles of quantum mechanics [4]. Nonetheless the Hawking effect, discovered nearly 30 years ago, is generally accepted as very credible and considered to be an essential ingredient of the as yet unknown correct theory of quantum gravity.

Recently Horowitz and Maldacena (HM) proposed a final-state boundary condition (FBC) [4] in an attempt to reconcile the unitarity of black hole evaporation with Hawking's semi-classical reasoning for massless scalar fields. The essence of the proposal is to impose a unique final-state boundary condition at the black hole singularity [4]. The final boundary state is an entangled state of the collapsing matter and infalling Hawking radiation [4-9]. When a black hole evaporates, particles are created in entangled pairs with one falling into the black hole and the other radiated to infinity. The projection of the final boundary state at the black hole singularity collapses the state into one associated with the collapsing matter and transfers the information to the outgoing Hawking radiation [4-9]. One of the critical assumptions in the HM proposal is that the internal quantum state of the black hole can be represented by entangled states of collapsing matter and infalling Hawking radiation. This conjecture is important because the final state boundary condition of the HM proposal is based on this entangled internal quantum state [4]. Recently, one of us proved that the HM conjecture is true in the special case of collapsing gravitational shell of the Schwarzschild black hole by considering a massless scalar field [8].

From a mathematical point of view, the Hawking effect arises from the logarithmic phase discontinuity of the wave functions [10]. Fermionic Hawking radiation is therefore also expected to exist (eg. neutrino radiation) [11-14], and has recently been shown to arise as a tunnelling effect [15]. It is therefore an interesting question as to whether the HM conjecture is also applicable to Dirac fields as well [7]. In this paper we consider this question and construct a modified FBC for massless neutrino fields for a shell that gravitationally collapses to a Schwarzschild black hole.

For s-wave collapse, a stationary Schwarzschild black hole has an effective (1+1) dimensional metric represented by [11]

$$ds^2 = -\left(1 - \frac{2M}{r}\right)dt^2 + \frac{dr^2}{\left(1 - \frac{2M}{r}\right)} \tag{1}$$

where $M$ is the mass of the black hole. At $r = 2M$, the Schwarzschild spacetime has an event horizon. Extending our formulation to a more general (3+1) dimensional setting is mathematically straightforward. The Levi-Civita connection coefficients $\Gamma^\lambda{}_{\mu\nu}$ can be calculated by the Lagrange method [16] and are given by

$$\Gamma^r{}_{tt} = \left(1 - \frac{2M}{r}\right)\frac{M}{r^2}\ , \ \Gamma^r{}_{rr} = -\frac{M/r^2}{\left(1 - \frac{2M}{r}\right)}, \ \text{and}\ \Gamma^t{}_{tr} = \frac{M/r^2}{\left(1 - \frac{2M}{r}\right)}. \tag{2}$$

The gamma matrices are defined by $\gamma_\mu = e^\alpha{}_\mu \bar{\gamma}_\alpha$ where $e^\alpha{}_\mu$ are tetrads and $\bar{\gamma}_\alpha$ are the gamma matrices for the inertial frame [11]. For example, we obtain

$$\gamma_t = \left(1 - \frac{2M}{r}\right)^{1/2} \bar{\gamma}_0 \ \text{and}\ \gamma_r = \left(1 - \frac{2M}{r}\right)^{-1/2} \bar{\gamma}_1. \tag{3}$$

The affine connection is then given by

$$\Gamma_\mu = -\frac{1}{4}\gamma_\nu\left(\partial_\mu \gamma^\nu + \Gamma^\nu{}_{\mu\lambda}\gamma^\lambda\right), \tag{4}$$

and the Dirac equation in the Schwarzschild spacetime is

$$\left[i\gamma^\mu(\partial_\mu + \Gamma_\mu) - m\right]\psi = 0, \tag{5}$$

or

$$i\frac{\partial}{\partial t}\psi = \omega(r)\left[-i\bar{\gamma}^0\bar{\gamma}^1 \omega^{1/2}(r)\frac{\partial}{\partial r}\left(\omega^{1/2}(r)\psi\right) + m\bar{\gamma}^0\psi\right], \tag{6}$$

where $m$ is the mass of the Dirac particle and $\omega(r) = |1 - 2M/r|^{1/2}$. For a neutrino field, we have $m = 0$ and an additional constraint

$$(1 + i\gamma^5)\psi = 0 \tag{7}$$

where $\gamma^5$ is given by [11]

$$\gamma^5 = \begin{pmatrix} 0 & 1 & 0 & 0 \\ -1 & 0 & 0 & 0 \\ 0 & 0 & 0 & -1 \\ 0 & 0 & 1 & 0 \end{pmatrix}.$$

The positive frequency normal mode solution of (6) coming out of the past horizon of the black hole is obtained as

$$\psi^-_\Omega = \omega^{-1/2}(r)\left[u_r(\Omega)\exp(-i\Omega u) + A^-_\Omega v_r(\Omega)\exp(-i\Omega v)\right], \tag{8}$$

where $u_r(\Omega)$, $v_r(\Omega)$ are 4 component spinors given by

$$u_r(\Omega) = \begin{pmatrix} 2\Omega \\ 0 \end{pmatrix} \otimes \begin{pmatrix} 1 \\ 0 \end{pmatrix}, \quad v_r(\Omega) = \begin{pmatrix} 0 \\ 2i\Omega \end{pmatrix} \otimes \begin{pmatrix} 1 \\ 0 \end{pmatrix},$$

and $A_\Omega^-$ is the fraction of the wave incoming from infinity. Since the Dirac field is massless one can effectively regard $u_r(\Omega)$, $v_r(\Omega)$ as 2 component spinors. We will set $A_\Omega^-$ to be zero when we match this solution with Kruskal components at $H^-$. The advanced and retarded coordinates $u$, v are given by

$$u = t - r^*, v = t + r^*, r^* = r + 2M \ln|r/2M - 1|. \tag{9}$$

In Kruskal coordinates, the Schwarzschild metric becomes

$$ds^2 = -2M \frac{\exp(-r/2M)}{r} d\bar{u} d\bar{v}, \tag{10}$$
$$\bar{u} = -4M \exp(-u/4M), \quad \bar{v} = 4M \exp(v/4M).$$

Since the Killing vector in Kruskal coordinate is given by $\partial/\partial \bar{u}$ on $H^-$, the (unnormalized) positive frequency solution is

$$\bar{\psi}_{\bar{\Omega}} = \bar{U}_{\bar{u}}(\bar{\Omega}) \exp(-i\bar{\Omega}\bar{u}), \tag{11}$$

where $\bar{U}_{\bar{u}}(\bar{\Omega}) = \dfrac{1}{\left(2M \dfrac{e^{-r/2M}}{r}\right)^{1/4}} \begin{pmatrix} 1 \\ 0 \end{pmatrix} \otimes \begin{pmatrix} 1 \\ 0 \end{pmatrix}.$

For a neutrino field, the helicity condition (7), implies that the spinors are of the form $(\phi_1 \quad i\phi_1 \quad \phi_2 \quad -i\phi_2)^\dagger$ (instead of $(\phi_1 \quad 0 \quad \phi_2 \quad 0)^\dagger$) and as a result the spinors in (8) and (11) should be modified accordingly.

Using $\exp(-i\Omega u) = (|\bar{u}|/4M)^{i4M\Omega}$ and $\exp(-i\Omega v) = (|\bar{v}|/4M)^{i4M\Omega}$, and the fact that $\bar{u} < 0$ in region I and $\bar{u} > 0$ in region II [8] (Fig. 1), the wave coming out of the past horizon of the black hole on $H^-$ can be written as

$$\psi_\Omega^- = \left(e^{2\pi M\Omega} {}_{out}\phi_\Omega + e^{-2\pi M\Omega} {}_{in}\phi_\Omega\right)/(2\cosh(4\pi M\Omega))^{1/2}, \tag{12}$$

where ${}_{out}\phi_\Omega$ vanishes inside the event horizon(region II) and ${}_{in}\phi_\Omega$ vanishes in the exterior region of the black hole (region I). Since $(|\bar{u}|)^{i4M\Omega}$ has a phase discontinuity of $e^{4\pi M\Omega}$ across the boundary surface $H^+$ at which $\bar{u}$ changes sign, the components ${}_{out}\phi_\Omega$ and ${}_{in}\phi_\Omega$ have the same phase discontinuity. We denote the normalized form of the Kruskal solution (11) as $\Psi_\Omega^-$ (as for the case of the scalar wave [17]). Here we have also used the fact that $\bar{v} = 0$ on $H^-$ and the Dirac inner product $(\psi_\Omega^-, \psi_\Omega^-) = \int dr (\psi_\Omega^-)^\dagger \bar{\gamma}_0 \gamma^0 \psi_\Omega^-$ [14] in the derivation. The above definition of the positive frequency solution leads to the Bogoliubov transformation [8] of the second-quantized fields and the ground state $|\Phi_0\rangle_{in \otimes out}$ that looks like the vacuum in the far past is the entangled state on the Fock space $\mathsf{H}_{in} \otimes \mathsf{H}_{out}$

$$|\Phi_0\rangle_{in \otimes out} = \frac{1}{(1+e^{-8\pi M\Omega})^{1/2}} \sum_{n=0,1} (-1)^n e^{-4n\pi M\Omega} |\bar{n}_\Omega\rangle_{in} \otimes |n_\Omega\rangle_{out}, \qquad (13)$$

where $|n_\Omega\rangle_\alpha$ is the second-quantized fermion number state for a mode $\Omega$ in a Fock space $\mathsf{H}_\alpha$ and $|\bar{n}_\Omega\rangle_\alpha$ is the corresponding anti-particle state.

We now show that the Dirac field inside the event horizon also can be decomposed into collapsing matter and the advanced wave incoming from infinity having similar form as the Hawking radiation. The Penrose diagram of a collapsing star is shown in Fig. 2 including the region II (black hole). The collapsing shell metric in two-dimension is given by [8,17]

$$ds^2 = \begin{cases} -d\tau^2 + dr^2, & r < R(\tau) \\ -\left(1-\frac{2M}{r}\right)dt^2 + \frac{dr^2}{1-\frac{2M}{r}}, & r > R(\tau), \end{cases} \qquad (14)$$

with the shell radius $R(\tau)$ defined by

$$R(\tau) = \begin{cases} R_o, & \tau < 0 \\ R_o - v\tau, & \tau > 0. \end{cases} \qquad (15)$$

We define the advanced and retarded null coordinates as

$$\begin{aligned} V &= \tau + r - R_o, \quad U = \tau - r + R_o, \\ v^* &= t + r - R_o^*, \quad u^* = t - r^* + R_o^*, \end{aligned} \qquad (16)$$

with $R_o^* = R_o + 2M\ln(R_o/2M - 1)$. The null coordinates are chosen such that the shell begins to collapse at $U = V = u^* = v^* = 0$ [9]. In these coordinates, the metric is given by

$$ds^2 = \begin{cases} -dUdV & \text{inside the shell,} \\ -(1-2M/r)du^*dv^* & \text{outside the shell.} \end{cases} \qquad (17)$$

For $r > R(\tau)$, the Levi-Civita connection coefficients and gamma matrices are given by

$$\begin{aligned} \Gamma^{v^*}{}_{v^*v^*} &= -\Gamma^{u^*}{}_{u^*u^*} = 2M/r^2, \\ \gamma_{v^*} &= 1/2(1-2M/r)^{1/2}(\bar{\gamma}_0 + \bar{\gamma}_1), \\ \gamma_{u^*} &= 1/2(1-2M/r)^{1/2}(\bar{\gamma}_0 - \bar{\gamma}_1), \\ \Gamma_{u^*} &= \Gamma_{v^*} = \frac{M}{2r^2}\bar{\gamma}^0\bar{\gamma}^1. \end{aligned} \qquad (18)$$

The Dirac equation for a massless neutrino becomes

$$\left[(\partial_{u*}+\partial_{v*})-\bar{\gamma}^0\bar{\gamma}^1(\partial_{u*}-\partial_{v*})+\frac{M}{r^2}\bar{\gamma}^0\bar{\gamma}^1\right]\psi=0, \quad (19)$$

with $(1+i\gamma^5)\psi=0$. The solution is given by

$$\psi=\omega^{-1/2}(r)\left[\bar{F}_{u*}(\Omega)\exp(-i\Omega u*)+\bar{G}_{v*}(\Omega)\exp(-i\Omega v*)\right], \quad (20)$$

where $\bar{F}_{u*}$ and $\bar{G}_{v*}$ are 4-component spinors

$$\bar{F}_{u*}(\Omega)=\begin{pmatrix}1\\i\\0\\0\end{pmatrix}, \quad \bar{G}_{v*}(\Omega)=\begin{pmatrix}0\\0\\1\\-i\end{pmatrix}. \quad (21)$$

For $r<\mathring{R}(\tau)$, the Dirac equation for a massless neutrino is given by

$$\left[\bar{\gamma}^0(\partial_U+\partial_V)-\bar{\gamma}^1(\partial_U-\partial_V)\right]\Psi=0 \quad (22)$$

with $(1+i\gamma^5)\Psi=0$. The solution is of the form

$$\Psi=\begin{pmatrix}f(U)\\if(U)\\g(V)\\-ig(V)\end{pmatrix}, \quad (23)$$

where $f$ and $g$ are arbitrary functions. The solutions (20) and (23) are connected by the following relations at the shell boundary [8]

$$u*=-4M\ln\left(1-\frac{vU}{(1+v)(R_0-2M)}\right), \quad v*=4M\ln\left(1-\frac{vV}{(1-v)(R_0-2M)}\right). \quad (24)$$

The massless neutrino field inside the black hole coming from infinity is given by

$$\psi_\Omega^+=\omega^{-1/2}(r)\left[v_r(\Omega)\exp(-i\Omega v)+A_\Omega^+ u_r(\Omega)\exp(-i\Omega u)\right] \quad (25)$$

The normal mode on $H^+$ (Fig. 2) becomes

$$\bar{\psi}_\Omega^+=\omega^{-1/2}(r)\bar{V}(\Omega)e^{i\Omega R_0*}\left|1-\frac{vW}{(1-v)(R_0-2M)}\right|^{-i4M\Omega}$$

$$=\left(e^{2\pi M\Omega}{}_M\phi_\Omega+e^{-2\pi M\Omega}{}_{in}\phi_\Omega\right)/(2\cosh(4\pi M\Omega))^{1/2}, \quad (26)$$

where ${}_M\phi_\Omega$ vanishes outside the shell, $V>(1-v)(R_o-2M)/v$ and ${}_{in}\phi_\Omega$ vanishes inside the shell, $V<(1-v)(R_o-2M)/v$. The above positive frequency solution implies that the ground state $|\Phi_0\rangle_{M\otimes in}$ in second-quantized fields is also an entangled state on $\mathsf{H}_M\otimes\mathsf{H}_{in}$ and is given by

$$|\Phi_0\rangle_{M\otimes in} = \frac{1}{(1+e^{-8\pi M\Omega})^{1/2}} \sum_{n=0,1} (-1)^n e^{-4n\pi M\Omega} |n_\Omega\rangle_M \otimes |\bar{n}_\Omega\rangle_{in} . \tag{27}$$

We propose this as a modified FBC for massless spinor fields. We shall now investigate its consequences.

We assume that the initial quantum state of the black hole belongs to a two dimensional Hilbert space $H_M$ and $|\psi\rangle_M$ is the initial quantum state of the collapsing matter. Let's assume that the orthonormal bases for $H_M$ are $\{|0\rangle_M, |1\rangle_M\}$. In the following, we consider only the single mode and drop the subscript $\Omega$. The FBC requires a maximally entangled quantum state in $H_M \otimes H_{in}$, which is called the final boundary state and is given by

$$\begin{aligned}{}_{M\otimes in}\langle\Psi| &= \sum_{n=0,1} A_n^* {}_M\langle n| \otimes {}_{in}\langle\bar{n}|(S_M \otimes I) , \\ A_n &= (1+\exp(-8\pi M\Omega))^{-1/2}(-1)^n \exp(-4n\pi M\Omega) .\end{aligned} \tag{28}$$

where $S_M$ is a random unitary transformation. The initial matter state is given by $|\psi\rangle_M = c_o |0\rangle_M + c_1 |1\rangle_M$. Then the initial matter state $|\psi\rangle_M$ evolves into a state in $H_M \otimes H_{in} \otimes H_{out}$, which is denoted by $|\Psi_0\rangle_{M\otimes in\otimes out} = |\psi\rangle_M \otimes |\Phi_0\rangle_{in\otimes out}$. The transformation from the quantum state of collapsing matter to the state of outgoing Hawking radiation is given by the following final state projection

$$\begin{aligned}|\phi_0\rangle_{out} &= {}_{M\otimes in}\langle\Psi||\Psi_0\rangle_{M\otimes in\otimes out} \\ &= |A_0|^2 \left(c_0 {}_M\langle 0|S_M|0\rangle_M + c_1 {}_M\langle 0|S_M|1\rangle_M\right)|0\rangle_{out} + |A_1|^2 \left(c_0 {}_M\langle 1|S_M|0\rangle_M + c_1 {}_M\langle 1|S_M|1\rangle_M\right)|1\rangle_{out} \\ &= |A_0|^2 \left(c_0 {}_{out}\langle 0|S_{out}|0\rangle_{out} + c_1 {}_{out}\langle 0|S_{out}|1\rangle_{out}\right)|0\rangle_{out} + |A_1|^2 \left(c_0 {}_{out}\langle 1|S_{out}|0\rangle_{out} + c_1 {}_{out}\langle 1|S_{out}|1\rangle_{out}\right)|1\rangle_{out} \\ &= \left(|A_0|^2 |0\rangle_{out}\langle 0| + |A_1|^2 |1\rangle_{out}\langle 1|\right) S_{out} \left(c_0 |0\rangle_{out} + c_1 |1\rangle_{out}\right) \\ &= P S_{out} |\psi\rangle_{out} \end{aligned}$$
(29)

where ${}_M\langle j|S_M|i\rangle_M = {}_{out}\langle j|S_{out}|i\rangle_{out}$ defines the unitary transformation $S_{out}$ and where $P = |A_0|^2 |0\rangle_{out}\langle 0| + |A_1|^2 |1\rangle_{out}\langle 1|$ is a density operator ($|A_0|^2 + |A_1|^2 = 1$), which acts as a weighted measurement.

This shows that the pure states always evolve to pure states under the black hole evaporation irrespective of the boundary condition at the event horizon, thus indicating that the black hole information paradox can be resolved for the case of fermions if the initial matter state is pure state.

We now consider the case when the initial matter state is mixed state. The mixed state is represented by the density operator $\rho_M = \sum_{n=0,1} C_n |n\rangle_M \langle n|$ with $\sum_{n=0,1} C_n = 1$. The information content of the mixed state is found to be dependant on the mixedness which is defined as $M_x = Tr(\rho_M{}^2)$ [18]. In order to get the final density operator at evaporation, we need to calculate ${}_{M\otimes in}\langle \Psi | [\rho_M \otimes |\Phi_0\rangle_{in\otimes out} \langle \Phi_0|] |\Psi\rangle_{M\otimes in}$, which is given by

$$\begin{aligned}
&{}_{M\otimes in}\langle \Psi | [\rho_M \otimes |\Phi_0\rangle_{in\otimes out} \langle \Phi_0|] |\Psi\rangle_{M\otimes in} \\
&= \sum_{i,j,k=0,1} C_i |A_j|^2 |A_k|^2 \left( {}_M\langle j|S_M|i\rangle_M \langle i|S_M{}^\dagger|k\rangle_M \right) |j\rangle_{out}\langle k| \\
&= \sum_{i,j,k=0,1} C_i |A_j|^2 |A_k|^2 \left( {}_{out}\langle j|S_{out}|i\rangle_{out} \langle i|S_{out}{}^\dagger|k\rangle_{out} \right) |j\rangle_{out}\langle k| \\
&= \left( \sum_{j=0,1} |A_j|^2 |j\rangle_{out}\langle j| \right) S_{out} \left( \sum_{i=0,1} C_i |i\rangle_{out}\langle i| \right) S_{out}{}^\dagger \left( \sum_{k=0,1} |A_k|^2 |k\rangle_{out}\langle k| \right) \\
&= D S_{out} \rho_{out} S_{out}{}^\dagger D,
\end{aligned} \quad (30)$$

where $D = \sum_{n=0,1} |A_n|^2 |n\rangle\langle n|$ is the distortion operator. Further more, $Tr(D) = 1$ and $Tr(D^2) \leq 1$. The final density operator is obtained by normalizing eq. (30):

$$\rho_f = \frac{D S_{out} \rho_{out} S_{out}{}^\dagger D}{Tr(D S_{out} \rho_{out} S_{out}{}^\dagger D)}. \quad (31)$$

The measure of mixedness is calculated as

$$Tr(\rho_f{}^2) = Tr(W \rho_{out} W \rho_{out}), \quad (32)$$

where

$$W = \frac{S_{out} D^2 S_{out}{}^\dagger}{Tr(D S_{out} \rho_{out} S_{out}{}^\dagger D)}. \quad (33)$$

Then we obtain $Tr(\rho_f{}^2) \leq Tr(\rho_{out}{}^2) = Tr(\rho_M{}^2)$ since for all bounded linear operator $X$ and density operator $T$, we have the following inequality [19]

$$|Tr(XT)| \leq \|X\| \|T\|_1, \quad (34)$$

where $\|\cdot\|$ is an operator norm and $\|\cdot\|_1$ is a trace norm. This suggests that the measure of mixedness is expected to decrease under evaporation or the information content is to decease. A mixed state arises due to the initial matter state being entangled with other quantum systems when the black hole was about to form, and results from taking a partial trace over these other systems. So we can regard black hole evolution as unitary.

The final-state boundary condition for the neutrino fields resembles that of the bosons [4-9] so that the black hole evaporation can be represented by the unified final-state boundary conditions. It is interesting to note that the final state projections described by equations (28) and (29) resemble universal teleportation protocol [20, 21], where entanglement plus local measurement and unitary transformation enables teleportation. In this protocol, the black hole evaporation is analogous to the measurement procedure done by Alice. The major difference exists, however, between the black hole evaporation and the quantum teleportation. In the latter, Bob needs a complete measurement results obtained by the classical channel to reconstruct the quantum state. On the other hand, the former doesn't need a classical channel because the evaporation is equivalent to sending the random unitary transformation as well as the state itself.

We close by pointing out that the Hilbert space of the black hole will in general be a tensor product of the two-dimensional Hilbert space for each fermionic mode times some other (bosonic) Hilbert space corresponding to other matter. The full FBC will be the fermionic condition we proposed in this paper combined with an FBC for other infalling matter. The complete FBC will be of the form $\left(_{M \otimes in}\langle\Psi|\right)_{fermion} \otimes \left(_{M \otimes in}\langle\Psi|\right)_{boson}$ and so is the Unruh state.

**Acknowledgements** This work was supported by the Korea Science and Engineering Foundation, the Korean Ministry of Science and Technology through the Creative Research Initiatives Program R17-2007-010-01001-0(2007) and the Natural Sciences and Engineering Research Council of Canada. D. A. would like to thank M. S. Kim for the discussion about the measure of mixed states. We are grateful to the organizers of the APCTP Focus Program on New Frontiers in Black Hole Physics where discussions initiating this work took place.

**Figure legends**

**Fig. 1.** The Kruskal exptention of the Schwarzschild spacetime [22,23]. In region *I*, null asymptotes $H_+$ and $H_-$ act as futue and past event horizons, respectively. The boundary lines labelled $J^+$ and $J^-$ are fute and past null infinities, respectively, and $i^o$ is the spacelike infinity.

**Fig. 2.** Penrose diagram of a collapsing star [23]. The region *I* is a fragmentation of Fig. 1 including the region *II* (black hole). The null ray $\gamma$ passes through the center of the collapsing matter and emerges to form the event horizon $H^+$.

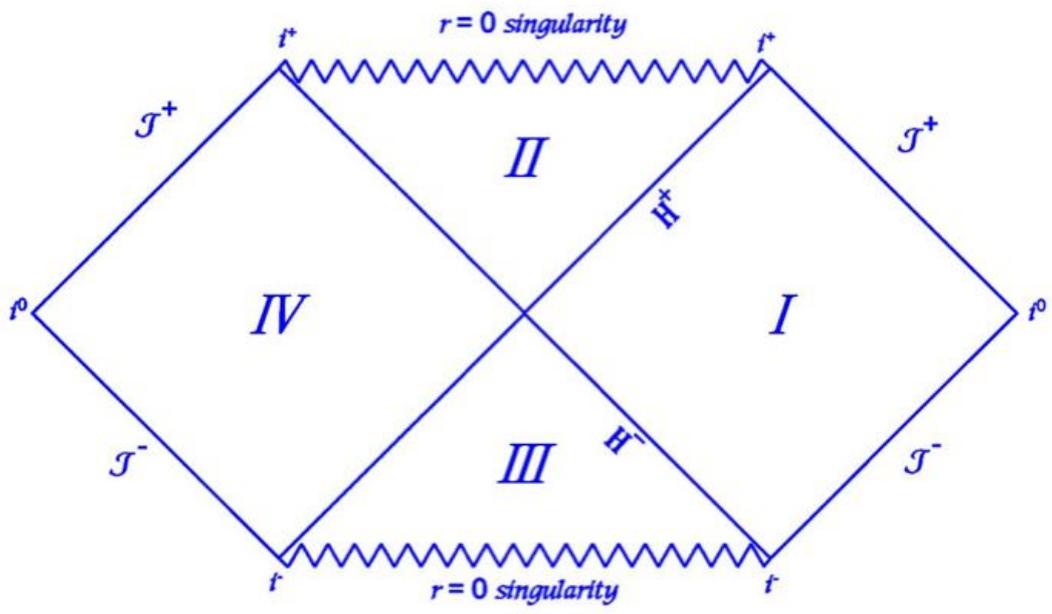

Fig. 1

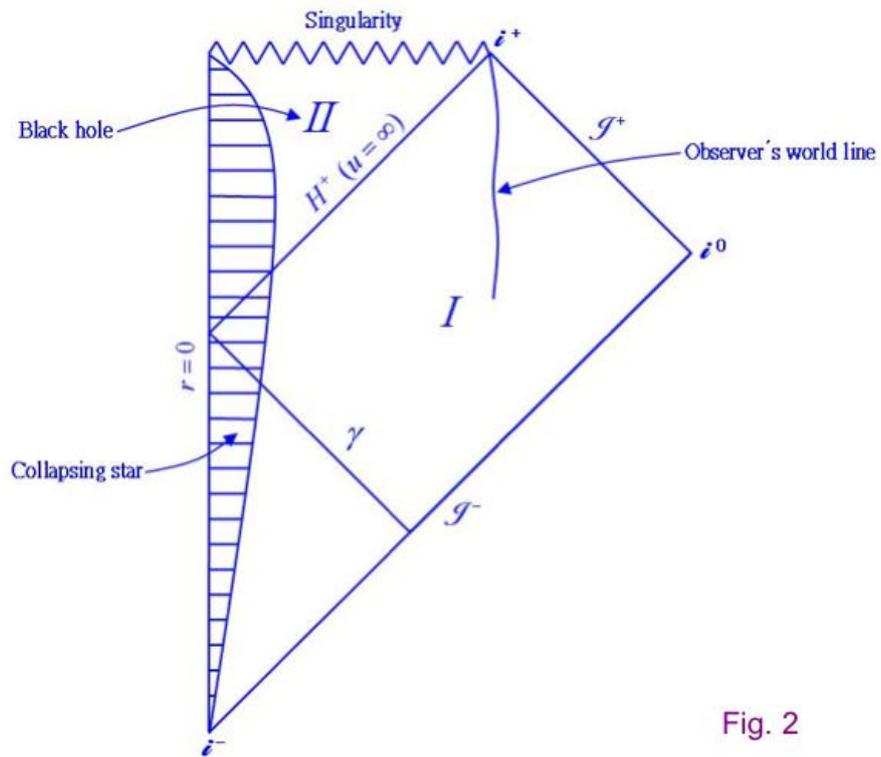

Fig. 2